\documentclass[12pt]{article}
\usepackage{geometry}
\usepackage{hyperref}
\usepackage{url}
\usepackage{latexsym}
\usepackage[centertags]{amsmath}
\usepackage{amssymb}
\usepackage{amsthm}
\usepackage{enumerate}
\usepackage{graphicx}
\usepackage{amsfonts,amssymb}
\usepackage{amsmath}

\usepackage[all]{xy}  
\usepackage{stmaryrd}

\usepackage{tikz}
\usepackage{tikz-cd} 
\usetikzlibrary{shapes.misc,arrows,decorations.markings}
\usetikzlibrary{matrix}

\newtheorem{Theorem}{Theorem}
\newtheorem{Definition}{Definition}
\newtheorem{Proposition}{Proposition}
\newtheorem{Lemma}{Lemma}

\newtheorem{Statement}{Statement}

\newcommand {\tr}{\mbox{${\mathrm{tr}}$}}

\newcommand{\DerC}{\mbox{${\mathrm{Der}(C)}$}}

\newcommand{\R}{\mathbb{R}}

\newcommand{\Cinf}{C^\infty }
\newcommand{\Minf}{{\Cinf}(M)}
\newcommand{\RAlg}{\R \textnormal{-} \mathbf{Alg}}
\newcommand{\RComAlg}{\R \textnormal{-} \mathbf{ComAlg}}

\begin{document}

\title{Einstein Algebras in a Categorical Context}

\author{Leszek Pysiak and Wies{\l}aw Sasin\\ \normalsize Institute of Mathematics and Cryptology, Military University of Technology\\
\normalsize Kaliskiego 2, 00-908 Warsaw, Poland \\[12pt]
Michael Heller and Tomasz Miller \\ \normalsize Copernicus Center for Interdisciplinary Studies, Jagiellonian University \\ 
\normalsize Szczepa\'{n}ska 1/5, 31-011 Cracow, Poland }

\date{\today}
\maketitle

\begin{abstract}
According to the basic idea of category theory, any Einstein algebra, essentially an algebraic formulation of general relativity, can be considered from the point of view of any object of the category of smooth algebras; such an object is then called a stage. If we contemplate a given Einstein algebra from the point of view of the stage, which we choose to be an ``algebra with infinitesimals'' (Weil algebra), then we can suppose it penetrates a submicroscopic level, on which quantum gravity might function. We apply Vinogradov's notion of geometricity (adapted to this situation), and show that the corresponding algebra is geometric, but then the infinitesimal level is unobservable from the macro-level. However, the situation can change if a given algebra is noncommutative. An analogous situation occurs when as stages, instead of Weil algebras, we take many other smooth algebras, for example those that describe spaces in which with ordinary points coexsist ``parametrized points'', for example closed curves (loops). We also discuss some other consequences of putting Einstein algebras into the conceptual environment of category theory.
\end{abstract}

\section{Introduction}
\label{sec::1}
Einstein's crowning achievement was the geometrisation of gravity. In the general theory of relativity, space-time is not just a stage on which physical processes take place, but it has become an active actor of the whole drama. Gravity has been integrated into the geometrical structure of space-time. The modern approach to geometry exploits the duality between space and the algebra of functions on it. Building geometry in terms of function algebras turns out to be more flexible and more amenable to generalisations than direct exploration of space by traditional methods. It was Vinogradov who defined what it means for an algebra to be geometric and studied this concept in depth \cite{Nestruev,Zoo}. In order to benefit from his work, we have to use an algebraic approach to general relativity. This had already been prepared by Geroch \cite{Geroch72} and two of the authors of this paper, who proposed and developed the notion of Einstein algebras \cite{Heller92,HS95a}. However, in order to derive new information from these already well-known theoretical facts, one should place Einstein algebras in a categorical context, i.e., to regard them as objects of the category \textbf{Einst} of Einstein algebras (which is a subcategory of the category $\Cinf $ of smooth algebras\footnote{For more generality we will also put our analyses in the context of $\R $-algebras.}). According to the basic idea of category theory, any Einstein algebra can be considered from the point of view of any object of the category ${\Cinf}$, which is then called a stage. Our leading example is when the stage is a Weil algebra, which can be interpreted as an ``Einstein algebra with infinitesimals''.

A given Einstein algebra, considered as an object of the category \textbf{Einst} or $\Cinf $, describes macroscopic gravity (understood as the geometry of the spectrum of this algebra), whereas if we contemplate this Einstein algebra from the point of view of the stage, which we choose to be the ``algebra with infinitesimals'' (Weil algebra), then we can suppose it penetrates the microscopic (or submicroscopic) level, on which quantum gravity functions. It turns out that if we apply Vinogradov's notion of geometricity (adapted to this situation), this level (i.e. quantum gravity level) is geometric, but then it is unobservable from the macro-level. Technically, both algebras, the algebra responsible for the macro-level and the algebra responsible for the micro-level, are isomorphic, so the micro-level does not add anything new to the macro-level. One way of breaking this isomorphism is by dropping the geometricity condition at the micro-level. This can be done, for instance, by relaxing the assumption that the considered algebra must be commutative. 

An analogous situation occurs when as stages, instead of Weil algebras, we take many other algebras, for example those that describe spaces in which with ordinary points there are associated ``generalized points'', for example closed curves (loops).

Putting general relativity, formulated in terms of Einstein algebras, in the context of category theory reveals those aspects of general relativity that remain invisible for standard tools. This works especially clearly when a given Einstein algebra, understood as an object of the relevant category, is considered from the point of view of other objects of this category. Then morphisms between these objects become dominant over objects themselves and disclose new relational properties of the system.

The plot of our work unfolds as follows. In Section \ref{sec::2}, we present some technical aspects of the algebraic approach to geometry, with particular emphasis on the notion of geometricity. We also take a step, in Section \ref{sec::3}, towards extending this notion to noncommutative algebras. From the outset we place our considerations in the context of category theory. In Section \ref{sec::4}, we define the category of Lorentz modules, and in Section \ref{sec::5} we discuss the category of Einstein algebras. In Subsection \ref{sec::5_1}, we define this category, and in Subsection \ref{sec::5_2}, we discuss its functorial aspects and show that the category of Einstein algebras is geometric\footnote{Strictly speaking, there are more than one categories of Einstein algebras, but they all have similar functorial and geometric properties.}. In the last two sections, we summarise the conclusions that emerge from our analyses. The first conclusion, commented upon in Section \ref{sec::6}, concerns the future theory of quantum gravity: if it is to be sought within the formalism analysed in this paper, and if it is to lead to effects measurable at the macro-level, it cannot be geometric (all ``background free'' theories are not geometric).  The second type of conclusions, considered in Section \ref{sec::7}, concerns rather philosophical aspects of general relativity: its placement in the milieu of category theory reveals new aspects of its relational character.

This paper can be regarded as a follow-up to \cite{Reports20}.

\section{Geometric algebras}
\label{sec::2}
Let $C$ be a commutative unital $\R $-algebra. Such algebras as objects, together with $\R $-linear algebra homomorphisms as morphisms, form a category of commutative $\R $-algebras, denoted $\RComAlg $. In this section, we will stay in the area of this category.

Let us define the real spectrum of an $\R $-algebra $C$ as
\begin{align*}
|C|_{\R } := [C, \R]_{\RComAlg}.
\end{align*}
Its point $\chi \in |C|_{\R }$ is a morphism $\chi: C \to \R $  
such that
\begin{align*}
& \chi (kf+g) = k\chi(f) + \chi (g),
\\
& \chi(fg) = \chi(f) \chi(g),
\\
& \chi ({\bf 1}_C) = 1
\end{align*}
for $k \in \R ,\, f, g \in C$.   

Now, for every $f \in C$, we define the function $\bar{f}^{\R }: |C|_{\R } \to \R$ by
\begin{align*}
\bar{f}^{\R } (\chi ) = \chi (f).
\end{align*}
If we define $\bar{C}^{\R } = \{\bar{f}^{\R } \, | \, f \in C \}$, we obtain the ringed space
$(|C|_{\R }, \bar{C}^{\R })$.

Let us consider the $\R $-algebra homomorphism $\Theta : C \to \bar{C}^{\R }$ given by $\Theta(f) = \bar{f}^{\R }$. It is obviously a surjection but not necessarily an injection. Let us define $J(C) = \bigcap_{\chi \in |C|_{\R }} \mathrm{ker}\, \chi $.

\begin{Lemma}
\label{lem1}
\begin{align*}
\mathrm{ker} \, \Theta = J(C).
\end{align*}
\end{Lemma}

\noindent \textbf{Proof.} We have the following chain of equivalences:
\begin{align*}
f \in \mathrm{ker}\, \Theta & \ \ \Leftrightarrow \ \ \Theta (f) = 0 \ \ \Leftrightarrow \ \ \bar{f}^{\R }(\chi ) = \chi (f) = 0, \forall \chi \in |C|_{\R }  
\\
& \ \ \Leftrightarrow \ \ f \in \mathrm{ker} \, \chi, \forall \chi \in |C|_{\R } \ \ \Leftrightarrow \ \ f \in \bigcap _{\chi \in |C|_{\R }} \mathrm{ker} \, \chi.
\end{align*}
\hfill $\Box$

Following Vinogradov \cite{Nestruev,Zoo} we adopt the following definition
\begin{Definition}
An $\R $-algebra $C$ is said to be geometric if the following two equivalent conditions are satisfied
\begin{align*}
(i) & \quad J(C) = 0,
\\
(ii) & \quad \Theta \; \mathrm{is \; an \; isomorphism}.
\end{align*}
\end{Definition}

Vinogradov \cite{Zoo} observes that the elements of $\ker \Theta $ are in a sense invisible and calls them ``ghosts'', i.e., an algebra is geometric if it has no ghosts. The name ``geometric'' is justified in so far as an algebra with this property can be interpreted as a function algebra on its spectrum.

Our next move is to generalise the above construction by looking at a commutative $\R$-algebra $C$ from the perspective of different stages (different $\R $\textnormal{-}algebras) what is a typical procedure in category theory. Let us then consider the functor:
\begin{align*}
[C,-]_{\RComAlg }: \RComAlg \to \mathbf{Set}.
\end{align*}

If $A$ is an object of the category $\RComAlg $, we call $|C|_A = [C,A]_{\RAlg }$ the $A$-spectrum of $C$, and the arrows $\rho : C \to A$ the points of $|C|_A$. We then proceed in close analogy to the above case when $\R $ was a stage.

For any $f \in C$ we thus define $\bar{f}^A: |C|_A \to A$ by 
\begin{align*}
\bar{f}^A (\rho ) = \rho (f), \quad \rho: C \to A
\end{align*}
and denote $\bar{C}^A =\{\bar{f}^A \, | \, f \in C\}$. We thus have a (generalised) ringed space $(|C|_A, \bar{C}^A)$. Let us consider the homomorphism $\Theta^A : C \to \bar{C}^A$ given by 
\begin{align*}
\Theta^A(f) = \bar{f}^A.
\end{align*}
It is obviously a surjection. Just as above, there is the ideal
\begin{align*}
J^A(C) = \bigcap_{\rho \in |C|_A} \mathrm{ker}\, \rho,
\end{align*}
and we also have the following analogue of Lemma \ref{lem1}.
\begin{Lemma}
\label{lem2}
\begin{align*}
\ker \Theta^A = J^A(C).
\end{align*}
\end{Lemma}

This enables us to formulate the following definition.
\begin{Definition}
An $\R $-algebra $C$ is said to be $A$-geometric if the following two equivalent conditions are satisfied
\begin{align*}
(i) & \quad \ker \Theta^A = \bigcap_{\rho \in |C|_A} \mathrm{ker}\, \rho = 0,
\\
(ii) & \quad  \Theta^A \; \mathrm{is \; an \; isomorphism}.
\end{align*}
\end{Definition}

If it is necessary to distinguish the case when the stage is $\R $ or $A$, we will refer to a given algebra as $\R$-geometric or $A$-geometric, respectively.

\begin{Lemma}
\label{lem3}
If $\R $-algebra $C$ is $\R $-geometric then $C$ is $A$-geometric, for every $A \in \RComAlg $. 
\end{Lemma}

\noindent \textit{Proof.} Let $f \in C$. We have: 
\begin{align*}
\bar{f}^A = 0 \ \ \Leftrightarrow \ \ \rho (f) = 0 \ \mathrm{ for \; every} \; \rho \in |C|_A.
\end{align*}
In particular, for $\rho = \chi \cdot \mathbf{1}_A, \, \chi \in |C|_{\R }$,
\begin{align*}
\bar{f}^A(\chi \cdot \mathbf{1}_A) = 0 \ \ \Leftrightarrow \ \ \chi (f) \cdot \mathbf{1}_A =0 \ \ \Rightarrow \ \ \chi (f), \; \forall \chi \in |C|_{\R }.
\end{align*}
Hence, since $C$ is ${\R }$-geometric, $f = 0$. \hfill $\Box $

The lemma is obviously also true ``in the opposite direction''. It states that it is sufficient to check $\R $-geometricity to conclude geometricity at any stage.

The ``quality'' of the considered spaces can be significantly improved if we restrict our analyses from the category of $\R $-algebras to its subcategory of smooth algebras (and their smooth homomorphisms as morphisms), denoted by $\Cinf $. Of course, all our results will be preserved, but in many respects they could be significantly "smoothed out". 

Let then now $C \in \Cinf $, and let us consider its spectrum $|C|_A := [C, A]_{\Cinf }$ with $A \in \Cinf $ (using the same symbol for the spectrum as in the case of the broader category $\RComAlg $ should not lead to confusion). By definition, any $\rho \in |C|_A$ satisfies
\begin{align*}
\rho (\omega (f_1, \ldots , f_n)) = \omega(\rho(f_1), \dots , \rho (f_n))
\end{align*}
for every $\omega \in \Cinf (\R^n)$ (cf. \cite{Reports20}). By repeating the constructions from the first part of this section, we obtain the ringed space $(|C|_A, \bar{C}^A)$, where $\bar{C}^A$ is itself a $\Cinf$-algebra with $\Theta^A : C \to \bar{C}^A$ being a $\Cinf$-algebra isomoprhism. In particular, we have that
\begin{align*}
\omega(\bar{f}^A_1, \ldots , \bar{f}^A_n) = \overline {\omega (f_1, \dots , f_n)}^A.
\end{align*}

Let us recall that a ringed space $(N,D)$, where $D$ is a set of real-valued functions on $N$ satisfying the following conditions
\begin{enumerate}
\item[(a)] 
$\textrm{sc} \, D=D$, where $\textrm{sc} \, D := \{\omega(f_1,...,f_n) \, | \, f_1,...,f_n\in D, \omega \in C^\infty({\mathbb{R}}^n)\}$,
\item[(b)]
$D_N=D$, where $D_N$ is the set of all local $D$-functions in the topology $\tau_D$ induced on $N$ by $D$, i.e. the weakest topology on $N$ in which all functions from $D$ are continuous,
\end{enumerate}
is called a \emph{Sikorski differential space}, and $D$ is called its \emph{structural algebra} \cite{Sikorski1967,Sikorski1971}.

Observe that the ringed space $(|C|_\R, \bar{C}^\R)$ constructed above automatically satisfies Sikorski's condition (a). If one endows $|C|_\R$ with the induced topology $\tau_{\bar{C}^\R}$, one can convert this ringed space into an actual Sikorski differential space by performing the ``localisation closure'' on $\bar{C}^\R$ \cite{Sikorski1967,Sikorski1971}. Note, however, that the localisation closure $(\bar{C}^\R)_{|C|_\R}$ is often larger than $\bar{C}^\R$ and so it is no longer isomorphic with $C$. All in all, we have the following fact.

\begin{Proposition}
Any geometric $\R$-algebra $C$ can be canonically embedded in a structural algebra of a Sikorski differential space.
\end{Proposition}
\noindent \textbf{Proof.} $\Theta : C \to \bar{C}^{\R }$ is an isomorphism and $\bar{C}^{\R }$ is canonically embedded in $(\textnormal{sc} \,\bar{C}^{\R })_{|C|_\R}$, a structural algebra of the Sikorski space $(|C|_\R, (\textnormal{sc} \,\bar{C}^{\R })_{|C|_\R})$. \hfill $\Box $

\vspace{0.5cm}
\noindent \textbf{Example: Weil algebras.} To close this section, let us consider the Weil algebra $W = \R \oplus \R \epsilon , \, \epsilon^2 = 0$ and ask if it is geometric. Any element of its spectrum $\pi \in |W|_{\R }  = [W, \R ]_{\Cinf }$ is given by $\pi (x + y\epsilon ) = x$. 

Indeed, we have $\pi (1) \cdot \pi (1) = \pi (1 \cdot 1) = 1$, which means that either $\pi (1) = 1$ or $\pi (1) = - 1$, but the latter is excluded. Similarly, $\pi (\epsilon ) \cdot \pi (\epsilon ) = \pi (\epsilon^2) = 0$, and hence $ \pi (\epsilon ) =  0$. By linearity, we obtain that 
\begin{align*}
\pi (x + y \epsilon ) = x \pi (1) + y \pi (\epsilon ) = x.
\end{align*}

Hence, $|W|_{\R } = \{\pi \}$. The differential structure on a one-point spectrum is $\R $, i.e. $\bar{W}^{\R } = \R $ which is given by
\begin{align*}
\overline{x +y \epsilon} (\pi ) = \pi (x + y \epsilon ) = x \in \R ,
\end{align*}
and so $\bar{W}^{\R }$ is not isomorphic with $W$ (we have the injection $W \hookrightarrow \bar{W}^{\R }$ but it is not surjective). Therefore $W = \R \oplus \R \epsilon $ is not geometric. The same can be easily shown for $W^k = \R [\epsilon]$ with $\epsilon^{k+1} = 0$, for any fixed $k > 0$.

\section{Geometric algebras in a generalized sense}
\label{sec::3}
Vinogradov's concept of geometric algebra is limited to commutative algebras. This narrows the scope of the concept's application to physics, especially on the quantum and subquantum scales, where noncommutativity is the rule rather than the exception. In this section, we are taking a step towards extending the concept of geometricity to include at least some noncommutative situations.

Let $A$ be a noncommutative $\mathbb{R}$-algebra, and let $\mathcal{Z}(A)$ be its center.  The idea is to call the algebra $A$ geometric in a generalized sense if it is $\mathcal{Z}(A)$-geometric. To show that the concept is far from being obvious, let us consider the example of Grassmann algebras in which noncommutativity is only ``partial'' (these algebras have only antisymmetric ``part'').
 
Let us first recall the definition. Associative unitary algebra is said to be a Grassmann algebra if it has a system of linearly independent generators $(\xi_1, \ldots , \xi_q)$ such that 
\begin{align*}
\xi_i \xi_j + \xi_j \xi_i = 0,
\end{align*}
in particular $\xi_i^2 = 0$. Grassmann algebras will be denoted by $\Lambda $ or $\Lambda (\xi ) = \Lambda (\xi_i, \ldots , \xi_q )$ if generators, called canonial generators, are to be displayed \cite{Berezin,DeWitt,Rogers}. 

Any element $a \in \Lambda (\xi )$ can be written as
\begin{align*}
a = a(\xi ) = \sum_{k \geq 0} \sum_{i_1, \ldots , i_k} a_{i_1, \ldots , i_k} \xi_{i_1} \ldots  \xi_{i_k}.
\end{align*}

In order for this formula to be unequivocal, we must assume that the factors $a_{i_1, \ldots , i_k}$ are antisymmetric with respect to their indices or that the indices are ordered in the form $i_1 < i_2 < \ldots < i_k$. The element corresponding to $k = 0$ is proportional to the unit.

Grassmann algebras are $\mathbb{Z}_2$-graded, i.e. two subspaces of $\Lambda $, $\Lambda_0$ and $\Lambda_1$, can be chosen in such a way that
$\Lambda = \Lambda_0 \oplus \Lambda_1$ and $\Lambda_i \Lambda_j \subseteq \Lambda_{i+j}$, where addition of indices is understood modulo 2.
Elements of $\Lambda_0$ are called even, and elements of $\Lambda_1$ are called odd. Elements that are either even or odd are called homogeneous. $\Lambda_0$ is a subalgebra, but $\Lambda_1$ is only a vector subspace of $\Lambda $. Homogeneous elements $a, b \in \Lambda (\xi )$ satisfy the following ``supercommutativity'' relation
\begin{align*}
ab = (-1)^{\alpha(a) \alpha(b)}ba
\end{align*}
where $\alpha (a)$ is the so-called parity function defined to be equal 0 if $a$ is even, and 1 if $a$ is odd.

Let $C^\infty(M, \Lambda)$ be the algebra of all smooth functions on a differential manifold $M$ with values in $\Lambda $. If $(\xi_1, \ldots , \xi_q)$ is a system of canonical generators then any  $f(x, \xi) \in C^\infty(M, \Lambda )$, $x \in M$, can be written as
\begin{align*}
f(x, \xi ) = \sum_{k \geq 0} \sum_{i_1, \ldots , i_k} f_{i_1, \ldots , i_k}(x) \xi_{i_1} \ldots  \xi_{i_k}
\end{align*}
where $f_{i_1, \ldots , i_k} \in \Minf $.

The concepts of even and odd elements are naturally transferred to $C^\infty(M, \Lambda )$. The set of all even elements in $C^\infty(M, \Lambda)$ coincides with $C^\infty(M, \Lambda_0)$, and the set of all odd elements of $C^\infty(M, \Lambda)$ coincides with $C^\infty(M, \Lambda_1)$. To denote them, we will keep the same symbols as above.

Let us notice that $\Lambda_0$ is the center of $\Lambda $. We thus can define the algebra $C^\infty(M, \Lambda )$ geometric in a generalized sense (in this case it could be called supergeometric) if it is $\Lambda_0$-geometric. However, the sets of morphisms with codomain in $\Lambda_0$ and of those with codomain in $\Lambda$ are not equal. The former set naturally embeds in the latter, but $|C^\infty(M,\Lambda)|_{\Lambda}$ is much richer than $|C^\infty(M,\Lambda)|_{\Lambda_0}$. An illustrative example is provided by the morphism $\chi_{v,a} \in [C^\infty(M,\Lambda),\Lambda]_{C^\infty}$ given by
\begin{align*}
\chi_{v,a}(f) := f(x,0) + v(f(\cdot,0))a
\end{align*}
for any fixed vector $v$ tangent to $M$ at $x$ and any \emph{nilpotent} $a \in \Lambda$ (for example, $a$ may be taken as one of the generators $\xi$). 

The situation certainly deserves more attention and will be addressed in a forthcoming work.

\section{Category of Lorentz modules}
\label{sec::4}
Let us first define the category \textbf{Mod} of modules. Its objects are pairs $(V, R)$, where $R$ is a ring (not necessarily commutative or unital) and $V$ is a (left) $R$-module (the pair $(V, R)$ will also be called an $R$-module). Its morphisms $\Phi: (V, R) \to (V', R')$ are pairs of mappings $\Phi = (F, \phi)$, where $\phi : R \to R'$ is a morphism of rings, and $F: V \to V'$
satisfies the condition
\begin{align*}
F(r_1v_1 + \ldots + r_kv_k) = \phi (r_1)F(v_1) + \ldots + \phi (r_k)F(v_k),
\end{align*}
for $r_1, \ldots , r_k \in R$, $v_1, \ldots , v_k \in V$, or, in a compact form,
\begin{align*}
& F(v_1 + v_2) = F(v_1)+ F(v_2),
\\
& F(rv) = \phi(r)F(v).
\end{align*}
In this way, we have the category of $R$-modules with $(V, R)$ as objects and $\Phi = (F, \phi )$ as morphisms.



Let us now assume that $V$ is a free $R$-module of rank $n+1$, i.e. that every basis of $V$ has $n+1$ elements. Let $g: V \times V \to R$ be a 2-linear, nondegenerate, symmetric mapping, which we shall call a scalar product on $V$. Such a scalar product is said to have the Lorentz signature if on $V$ there exists an orthonormal basis $W_0, W_1, \dots , W_n$ in which
\begin{align*}
g(W_i, W_j) =
\left\{
\begin{array}{rl}
{\bf 0}_R & \mbox{if $i\neq j$} \\
-{\bf 1}_R & \mbox{if $i=j=1, \dots, n$} \\
{\bf 1}_R & \mbox{if $i=j=0$}
\end{array}
\right.
\end{align*}
If $g$ has a Lorentz signature, it is said to be a \emph{Lorentz metric}. 

The triple $(V, R, g)$, where $g$ is a Lorentz metric, will be called a Lorentz module (we shall also write $(V, g)$ if the ring $R$ is obvious). Let us consider two such $R$-modules $(V, R, g)$ and $(V', R', g')$, and a morphism of $R$-modules $\Phi = (F, \phi): (V, R) \to (V', R')$ such that
\begin{align*}
g'(F(X), F(Y)) = g(X, Y)
\end{align*}
for $X, Y \in V$.\footnote{Let us notice that $g'(F(X), F(Y)) = (F^*g')(X, Y)$.} Then $\Phi $ is said to be a morphism of Lorentz modules $(V, R, g) \to (V', R', g')$.

Lorentz modules, as objects, with mappings $\Phi = (F, \phi)$ as morphisms naturally form the category of Lorentz modules, denoted by \textbf{Lor}. It is a subcategory of the category of $R$-modules.

The archetypical example of a Lorentz module is $(\textrm{Der}(C^\infty(M)),C^\infty(M), g)$, where $(M,g)$ is a Lorentz manifold and $\textrm{Der}(C^\infty(M))$ is the $C^\infty(M)$-module of derivations\footnote{Recall that a derivation of an $\R$-algebra $C$ is an $\R $-linear map $X: C \to C $ satisfying the Leibniz condition $X(fg) = X(f)g + fX(g)$, $f,g \in C$.} of the algebra $C^\infty(M)$. This can be generalised by replacing $C^\infty(M)$ with a $C^\infty$-algebra $\bar{C}$ understood either as $\bar{C}^{\R }$ or as $\bar{C}^A, \, A \in \Cinf $, along with replacing the metric $g$ with a 2-linear, nondegenerate, symmetric mapping $\bar{g}: \mathrm{Der}(\bar{C}) \times \mathrm{Der}(\bar{C}) \to \bar{C}$ with the Lorentz signature. In what follows, we shall sometimes denote such a Lorentz module by $(\bar{C}, \bar{g})$.

\section{Category of Einstein Algebras}
\label{sec::5}
\subsection{Einstein Algebras}
\label{sec::5_1}
In this subsection, we briefly summarise standard geometric constructions leading to Einstein's equations resp. Einstein algebras (for full account see, e. g., \cite{Heller92,Neill,Sachs}).

Let $C$ be a geometric $\Cinf $-algebra and the module of its derivations $V = \DerC $. We assume that it is a free module of rank $n$. Let further  $(V, C, g)$ be a Lorentz  module. We thus have a symmetric and nondegenerate 2-form $g: V \times V \to C$.

From its nondegeneracy it follows that, for $X \in V$, we have $\iota_X g \in V^*$ with
\begin{align*}
(\iota_X g)(Y) = g(X,Y).
\end{align*}
This leads to the isomorphism $\Psi_g: V \to V^*$ given by 
\begin{align*}
\Psi_g(X) = \iota_X g \in V^*.
\end{align*}

With the help of the Lorentz metric $g$ we define the Levi-Civita connection $\nabla $, i.e., a unique connection $\nabla : V \times V \to V$ such that
\begin{align*}
(i) & \quad [X,Y] = \nabla_XY - \nabla_YX,
\\
(ii) & \quad Xg(Y,Z) = g(\nabla_XY,Z) + g(Y, \nabla_XZ),
\end{align*}
for all $X, Y, Z \in V$. This connection allows us to determine all basic curvature structures necessary to define Einstein's equations. First of all, the curvature tensor $R: V \times V \times V \to V$
\begin{align*}
R(X, Y)Z = \nabla_X \nabla_Y Z - \nabla_Y \nabla_X Z - \nabla_{[X, Y]}Z,
\end{align*}
and the endomorphism $R_{XY}: V \to V$ given by
\begin{align*}
R_{XY}(U) = R(X,U)Y.
\end{align*}
Now, we define the Ricci tensor
\begin{align*}
\mathrm{Ric} (X, Y) = \tr\, R_{XY}.
\end{align*}

We also define the endomorphism $\mathcal{R}: V \to V$ by
\begin{align*}
\mathcal{R}(X) = \Psi_g^{-1}(\iota_X \mathrm{Ric}).
\end{align*}
It satisfies
\begin{align*}
g(\mathcal{R}(X), Y) = \mathrm{Ric}(X, Y).
\end{align*}
Finally, we define the scalar curvature
\begin{align*}
r = \tr \, \mathcal{R}.
\end{align*}

Now, we have  all structures needed to define the Einstein algebra \cite{Geroch72,Heller92,HS95a}.

\begin{Definition}
The Lorentz module $(V, C, g)$ is called an Einstein algebra, if Einstein's equations are satisfied, i.e. if on the Lorentz module $(V, C, g)$ the following conditions hold
\begin{align*}
(i) & \quad \mathrm{Ric} - \tfrac{1}{2}rg  + \Lambda g = 8 \pi T,
\\
\textrm{or} \quad (ii) & \quad \mathrm{Ric} = \Lambda g,
\end{align*}
where $\Lambda $ is the cosmological constant and $T$ a suitable energy--momentum tensor.
\end{Definition}

If the energy--momentum tensor is zero $(T = 0)$, equations (i) reduce to equations (ii), called the \emph{vacuum} Einstein's equations\footnote{With $\Lambda$ suitably rescaled.}. In some applications (for example, to calculate effects in a weak gravitational field), one assumes $\Lambda = 0$ (for a more detailed discussion of the above definition see \cite{Heller92}).

Let us notice that the category of Einstein algebras, denoted \textbf{Einst} or $\mathbf{Einst}_0$, depending on whether (i) or (ii) is satisfied, is a subcategory of the category of Lorentz modules.

Rosenstock, Barrett and Weatherall \cite{Rosen} have demonstrated that the categories of general relativity and Einstein algebras are dual with respect to each other, which implies that they are ``theoretically equivalent'' (equivalent as theories), however these authors limited themselves to the case when, using our terminology, the starting point for constructing Einstein algebras is a differential manifold. In our approach, it is, in general, a ringed space $(|C|_A, \bar{C}^A)$ (of which differential manifold is a special instant with $A = \R$ and $C = \Cinf(M)$). In such a case, Einstein algebras are genuine generalisations of general relativity. For example, a relativistic world model with quasiregular space-time singularities, which evidently violates the smooth manifold structure, can be organized into an Einstein algebra \cite{Heller92}.

\subsection{Functorial Einstein Algebras}
\label{sec::5_2}
In constructing Einstein's algebra at a stage $A$, one could explicitly use the isomorphism $\Theta^A: C \to \bar{C}^A$, but for the sake of illustration we will sketch briefly the main steps of this construction.
Let $C$ be, as above, a geometric $\Cinf $-algebra. In this section, we consider the algebra $\bar{C}^A$ with $A \in \Cinf $. Its elements are of the form $\bar{f}^A: |C|_A \to A$ given by $\bar{f}^A(\rho ) = \rho(f), \, \rho \in |C|_A := [C, A]_{\Cinf }$, $f \in C$. By Lemma \ref{lem2}, $\bar{C}^A$ is a geometric algebra.

From the isomorphism of the algebras $C$ and $\bar{C}^A$ it follows that  all tensors, corresponding to these algebras, are also respectively isomorphic (see \cite{Reports20}). Thus we may proceed in close analogy with what has been done in the previous subsection. We construct the module of derivations $\bar{V}^A = \mathrm{Der}(\bar{C}^A)$. If $X \in \DerC$ then $\bar{X}^A (\bar{f}^A) = \overline{X f}^A$, and the Lorentz metric $\bar{g}^A$ on it satisfies
\begin{align*}
\bar{g}^A(\bar{X}^A,\bar{Y}^A) = \overline{g(X,Y)}^A.
\end{align*}
We thus have a Lorentz module $(\bar{V}^A, \bar{C}^A, \bar{g}^A)$, and if we proceed just as in the preceding subsection, we arrive at the following definition
\begin{Definition}
The Lorentz module $(\bar{V}^A, \bar{C}^A, \bar{g}^A)$ at a stage $A$ is called an Einstein algebra at the stage $A$ if Einstein's equations, analogous to (i) or (ii), are satisfied.
\end{Definition}

The above constructions can be summarised in the form of the following statement.
\begin{Statement}
If the triple $(V, C, g)$ is an Einstein algebra then $(\bar{V}^A, \bar{C}^A, \bar{g}^A)$ is an Einstein algebra at the stage $A$.
\end{Statement}

Let now $\bar{C}$ be a functorial $\Cinf $-algebra, i.e. an algebra the elements of which are
\begin{align*}
\bar{f} = \{\bar{f}^A \, | \, f \in C\}_{A \in \Cinf }.
\end{align*}

Consequently, we consider the functorial module of derivations $\bar{V} = \{\bar{V}^A\}_{A \in \Cinf }$, and by proceeding as above, we arrive at the following definition
\begin{Definition}
The functorial Lorentz module $(\bar{V}, \bar{C}, \bar{g})$ is called a functorial Einstein algebra if Einstein's equations, analogous to (i) or (ii), are satisfied.
\end{Definition}

And again, we summarise the above construction in the following way.
\begin{Statement}
If the triple $(V, C, g)$ is an Einstein algebra, then $(\bar{V}, \bar{C}, \bar{g})$ is a functorial Einstein algebra.
\end{Statement}
In fact, we have a functor from the category of Einstein algebras to the category of functorial Einstein algebras.

We can summarise our analyses by formulating the following theorem.
\begin{Theorem}
All three Einstein algebras $(V, C, g)$, $(\bar{V}^A, \bar{C}^A, \bar{g}^A)$ and $(\bar{V}, \bar{C}, \bar{g})$, for any $A \in \Cinf $, where $C$ is a geometric $\Cinf $-algebra, are isomorphic with each other. 
\end{Theorem}

\noindent \textbf{Proof.} We shall show this for the first two algebras; extension to the third algebra is obvious.

We have the isomorphism: $(\Theta^A)^{-1}: \bar{C}^A \to C$ which induces the isomorphism of modules of derivations $V = \DerC $ and $\bar{V}^A = \mathrm{Der}(\bar{C}^A)$. The equality $\bar{g}^A (\bar{X}^A, \bar{Y}^A) = \overline{g(X, Y)}^A$, $X, Y \in V$, guarantees that the Lorentz modules $(V, C, g)$ and $(\bar{V}^A, \bar{C}^A, \bar{g}^A)$ are isomorphic. \hfill $\Box $

\section{Infinitesimals and parametrised points}
\label{sec::6}
One of the important tools, used in this work, was the strict definition of geometric algebra proposed by Vingradov. As expected, the general theory of relativity (in terms of Einstein's algebras) turned out to be geometric in this strict sense. An algebra is geometric if it is isomorphic to the algebra of real-valued functions defined over its spectrum. Then, in terms of these functions, geometry may be developed. Therefore, no geometric algebra leads to the ``background free'' theory. Meanwhile, it is quite commonly believed that the final theory of quantum gravity should be ``background free''. If so, the corresponding algebra should not be geometric. Let us consider the following ``toy model''.

We associate with a given Einstein algebra a Weil algebra representing infinitesimal quantities, and make the working assumption that the quantum theory of gravity (at the Planck level) uses infinitesimal quantities. Below, we will replace the infinitesimal quantities by other structures, but this will not change the situation significantly.

Let us consider a Weil algebra of the form $W = \R \oplus \R \epsilon , \, \epsilon^2 = 0$, like in Example at the end of Section \ref{sec::2}, and an Einstein algebra $(V, C, g)$. It immediately follows from Lemma \ref{lem3} that the algebra $C$ is $W$-geometric. However, to see this more clearly, let us trace the reasoning more closely.

The ``real spectrum'' of $C$ is $|C|_{\R } := [C, \R ]_{\Cinf }$. Its ``points'' are the arrows: $\chi : C \to \R$. We define, as usual, $\bar{f}^{\R }: |C|_{\R } \to \R$ by 
\begin{align*}
\bar{f}^{\R }(\chi ) = \chi (f), \ f \in C
\end{align*}
and
\begin{align*}
\bar{C}^{\R } = \{\bar{f}^{\R } \, | \, f \in C\}.
\end{align*}
Now, we include infinitesimals by looking at the Einstein algebra from the stage $W$. The ``infinitesimal spectrum'' is $|C|_{W} = [C, W]_{\Cinf }$, and its ``points'' are arrows $\rho : C \to W$
given by
\begin{align*}
\rho (f) = \chi (f) + v(f)\epsilon
\end{align*}
where $f \in C$, $\chi \in |C|_{\R }$ and $v \in \mathrm{Der}_{\chi }C$; $v$ can be regarded as a vector tangent to $|C|_\R$ at the ``point'' $\chi $ \cite{Reports20}. 

And finally, we have the isomorphism $\Theta^W: C \to \bar{C}^W$, which is equivalent to saying that $C$ is $W$-geometric.
Since it is an isomorphism, the effects from the infinitesimal level (gravitational micro-level), do not contribute anything new to the level described by Einstein algebras (gravitational macro-level), that is, the micro-level is not visible from the macro-level. If this diagnosis is accurate and if we want to have quantum gravity effects measurable at the macro-level, then a quantum theory of gravity should be sought by constructing non-geometric models. In such models, the spectrum might be ``too small'', i.e. the algebra of real-valued functions on this spectrum is isomorphic only to a proper subalgebra of the original algebra. In the extreme case, the spectrum could be even empty. However, by allowing noncommutative algebras, the situation can change (see Section \ref{sec::3}).

We have used the Weil algebra example mainly for illustrative reasons: we can imagine that infinitesimals are associated with the lowest (Planck) level of physics, but as a stage we can choose, in principle, any object of the $\Cinf$-algebra category (or even the $\R$-algebra category, depending on the needs). This gives us a freedom in choosing the objects we want to consider on the micro-level. This is particularly clear on the example of the so-called parametric points. 

We could start with abstract $\R $-algebras, as we have done in the previous sections, and progressively move to differential spaces, but for the sake of simplicity let us start right away by considering two differential spaces $(M, \Cinf(M))$, representing space-time manifold, and $(P, \Cinf(P))$ a differential space representing a typical ``parametrized point''. 

We define
\begin{align*}
M^P = \{\phi: P \to M \, | \, \phi^*(\Cinf(M)) \subset \Cinf(P)\}.
\end{align*}

If $f \in \Cinf(M)$ then its extension to $M^P$ is defined to be $\hat{f}: M^P \to \Cinf (P)$ with
\begin{align*}
\hat{f}(\phi) := \phi^\ast(f) = f \circ \phi
\end{align*}
for any $\phi \in M^P$, which leads to
\begin{align*}
\Cinf(M^P) := \{\hat{f} \, | \, f \in \Cinf(M) \}.
\end{align*}

Arrows $\phi: P \to M$ are called parametrised points of $M$. Elements of $M^P$ may be, for example, ``loops'' (if $P=S^1$), tori (if $P=S^1 \times S^1$), ``strings'' (if $P=\R$ or $P=S^1$) or ``branes'' (if $P=\R^n$ etc.) We assume that dim $M$ $>$ dim $P$. 

We thus obtain the differential space $(M^P, \Cinf(M^P))$.
 
Let us define
\begin{align*}
M_0^P := \{\phi_m \, | \, m \in M\} \quad \textnormal{with} \quad \phi_m(p) = m, \ \ \forall p \in P.
\end{align*}
Clearly, we have the bijection $M \cong M_0^P$ given by $m \leftrightarrow \phi_m$, and the differential structure on $M_0^P$ is $\Cinf(M^P)|M_0^P$.
Indeed, we have
\begin{align*}
\hat{f}(\phi_m)(p) = (f \circ \phi_m)(p) = f(\phi_m(p)) = f(m). 
\end{align*}

We can also easily see that the two differential structures $\Cinf (M)$ and $\Cinf (M^P)$ are isomorphic, the isomorphism being given by $f \mapsto \hat{f}$. The situation is summarised in the following commutative diagram

\begin{center} 
\begin{tikzcd}
{(M, C^{\infty }(M))} \arrow[rrr, "{(\Phi, \, \Phi^*)}"] \arrow[rrrd, "{(\Phi, \, \Phi_0^*)}"', shift right] &  &  & {(M^P, C^{\infty}(M^P))}                                             \\
                                                                                          &  &  & {(M^P_0, C^{\infty }(M^P_0))} \arrow[u, "{(\iota, \, \iota^*)}"', hook]
\end{tikzcd}
\end{center}
Here we have defined:
\begin{align*}
& \Phi: M \to M^P \quad \textnormal{by} \quad \Phi(m) = \phi_m,
\\
& \iota : M^P_0 \to M^P \quad \textnormal{by} \quad \iota (\phi_m) = \phi_m,
\\
& \Phi_0: M \to M^P_0 \quad \textnormal{by} \quad \Phi_0(m) = \phi_m,
\end{align*}
\noindent
for any $m \in M$, and their pullbacks, correspondingly.

The mapping $\Phi_0$ is a bijection, whereas $\iota$ and $\Phi$ are injections. The geometry on $M^P$ is the same as that on $M^P_0$ since the respective algebras of functions are isomorphic.

As we can see, adding parametric points to space-time does not significantly change the differential structure of space-time; consequently, it does not change its geometry (geometry is encoded in the differential structure of space-time, not in the collection of its points). Moreover, the whole procedure does not change the differential dimension of space-time.

\section{Comments and interpretation}
\label{sec::7}

Let us now turn to drawing some conclusions from the algebraic formulation of general relativity (as Einstein algebras) and from placing it in a categorical context. Geroch's motivation, when he introduced the notion of Einstein algebras, was to reduce the role of sharply defined events as points in a space-time manifold. In quantum physics, for instance, ``the influence of the measuring apparatus on the system being observed cannot, even in principle,  be made arbitrarily small'' \cite{Geroch72}. In his opinion, Einstein algebras, in which the underlying manifold plays practically no role, may offer a convenient starting point to ``smearing out of events'' in the future quantum gravity theory. A fairly natural motivation for considering Einstein algebras is provided by the idea that smooth functions on a manifold can be interpreted as scalar fields on space-time \cite{Demaretetal}; Chen and Fritz \cite{ChenFritz} have even attempted to show that other physical fields too can be interpreted algebraically. 

It is significant that Einstein algebras have interested philosophers of science more than physicists. For instance, John Earman claimed that the Einstein algebra approach could make precise Leibniz relational philosophy of space-time. That is why he called them Leibniz algebras \cite[p. 193]{Earman1989}. As a reaction, Rynasiewicz \cite{Rynasiewicz} argued that in the algebraic approach points can be recovered, and philosophically one gains nothing by dealing with $\Cinf (M)$ rather than with $M$. The discussion, in a sense, was closed by Weatherall \cite{Weath}, who defined the category \textbf{GR} of standard general relativity theory and the category \textbf{AE} of Einstein algebras and, citing \cite{Rosen}, showed that there is a functor $F: \mathbf{GR} \to \mathbf{AE}$ which ``forgets nothing'', i.e. points in \textbf{AE} exist just as they do in \textbf{GR}.\footnote{We should only remember that our treatment of Einstein algebras is more general than the treatment considered by the cited authors, who limit themselves to the category of smooth manifolds.} We can add from ourselves that this is so precisely because Einstein algebras are geometric: one can always reconstruct their spectra, and hence all the points. 

Putting Einstein algebras into the context of category theory brings yet another benefit: it shows the aspect of relational character of general relativity, which the cited authors did not pay attention to. To see this in its full light, we must turn to tools often used in category theory.

Yoneda Lemma, in its covariant version, states that for  every object $X$ in a (locally small) category $E$, and every functor $F: E \to \mathbf{Set} $, the set of natural transformations from $[X, - ]_E$ to $F$ is isomorphic with $FX$ \cite[Section 0.2.3]{TopCat}, i.e.,
\begin{align*}
\mathrm{Nat}([X, - ]_E, F) \cong FX.
\end{align*}
Let us apply this result to the category $\Cinf $ and put $F = [A, -]_{\Cinf }$, $A \in \Cinf $. We thus obtain
\begin{align*}
\mathrm{Nat}([X, - ]_{\Cinf }, [A, - ]_{\Cinf }) \cong [A, X]_{\Cinf }.
\end{align*}
Now, the direct corollary from the Yoneda Lemma is
\begin{align*}
X \cong A \; \; \; \mathrm{iff} \; \; \; [X, - ]_{\Cinf } \cong [A, - ]_{\Cinf }.
\end{align*}
One direction in proving this corollary is the consequence of the fact that the Yoneda embedding is a functor, and the other direction of the fact that this functor is faithful \cite{TopCat}.

The corollary tells us that any object $A$ can be viewed as a functor $[A,-]_{\Cinf }$ (or as a functor $[-,A]_{\Cinf }$ if we repeat the whole argument for the contravariant version of the Yoneda lemma). This means that any object is entirely determined by all outgoing or incoming arrows. In other words, any algebra of smooth functions (any object of $\Cinf $) is completely determined by its relationship with other objects.

When we identify a given object with all arrows going to or coming from it, new elements appear in this object that were not originally there. This is typical for category theory. Treating these new elements on equal footing with ``original objects'' often makes the theory more transparent and  logically compact.

In the view of the above, any (functorial) Einstein algebra is completely determined by its relationship to other (functorial) Einstein algebras. Intuitively speaking, this is another expression of a strongly relational character of general relativity (and its possible generalisations). Let us notice that this does not only apply to Einstein algebras. By virtue of the generality of the Yoneda lemma, the same may be said of any theory which can be expressed in the language of category theory. This shows the universality of the ``principle of relationality''.

If a solution to Einstein equations is regarded as a pair $(M, g)$ where, for simplicity, we assume that $M$ is a manifold, and $g$ is a Lorentz metric on $M$ (or the equivalence class of physically equivalent Lorentz metrics on $M$), satisfying Einstein equations, then it can be represented as an Einstein algebra, an object of of the category \textbf{Einst} or $\mathbf{Einst}_0$. Consequently, every solution of Einstein equations is entirely determined by all arrows emanating from this solution to all other solutions or coming from all other solutions to this solution. It is well known that solutions of any differential equation (and Einstein equations are far from being an exception), are not independent entities but constitute elements of highly structured ``solution space'' (see, for instance, \cite{Fischer}); it seems, however, that the ``categorical holism'' goes further in combining all Einstein algebras into a single category.

\end{document}